\documentclass{PoS}
\usepackage{verbatim}
\usepackage{epsfig}

\title{Charm Quarks and the QCD Equation of State}

\ShortTitle{Charm Quarks and the QCD Equation of State}

\author{\speaker{Michael Cheng} for the RBC-Bielefeld Collaboration\\
        Department of Physics, Columbia University, New York, NY 10027, USA\\
        E-mail: \email{michaelc@phys.columbia.edu}}

\abstract{
We present a study of the effect of charm quarks on the QCD equation of state
using partially-quenched p4 charm quarks on a dynamically generated 2+1 flavor
background, at zero chemical potential.  We show preliminary results for
the charm quark contribution to the energy density and pressure in the 
high temperature region ($T_c < T < 4 T_c$) and compare it to the free-field 
calculation.  The charm quark mass is determined by measuring the charmonium 
spectrum.
}

\FullConference{The XXV International Symposium on Lattice Field Theory\\
                 July 30 - August 4 2007\\
                 Regensburg, Germany}

\begin{document}

\section{Introduction}
The equation of state(EoS) of QCD, i.e. thermodynamic
quantities such as the pressure ($p$), energy density ($\epsilon$), or
entropy density ($s$) as a function of the temperature, is important in understanding
the high-temperature behavior of QCD.  The EoS is not only of theoretical
interest, but is directly applicable to the dynamics of the quark-gluon
plasma (QGP), whether in the context of interpreting the results of
heavy-ion experiments or modelling the behavior of hot, dense
matter in the early universe.   

Currently, some aspects of the EoS are accessible at high
temperatures via perturbation theory \cite{Vuorinen:2003fs}, 
or at low temperatures via the hadron resonance gas model (HRG)
 \cite{BraunMunzinger:2003zd}.  
However, a truly non-perturbative,
first-principles calculation can only be done using lattice QCD.

Recently, there have been several detailed studies of the QCD EoS using
improved actions and physical (or almost-physical) values for the light
and strange quark masses \cite{Bernard:2006nj, Aoki:2005vt, Cheng:2007tp}.
While these studies accurately represent the three lightest quark flavors,
they, like all previous calculations, neglect the effect of the charm quarks.  
However, recent work \cite{Laine:2006cp} based on perturbation theory 
indicates that the charm quark contribution may start to become significant at
temperatures $T \sim 350$ MeV, well within the temperature
range covered by these latest EoS calculations.

In this work, we present a partially quenched study of the charm quark
contribution to the QCD equation of state using p4fat3 fermions on 
a 2+1f gauge background in the temperature range $0.90 T_c < T < 4.2 T_c$.
Currently, the calculations are done only for $N_t =4$ lattices, with the
intention of continuing to $N_t = 6$ and possibly $N_t = 8$.

\section{Equation of State on the lattice}
We will first review the formalism for calculating the equation of state on
the lattice via the "integral method", as described in \cite{Bernard:1996cs}.
First, consider the grand canonical partition function of QCD at vanishing
quark chemical potential ($\mu_q = 0$):
\begin{equation}
\label{eqn:partition}
Z(V,T)  =  \int[\mathcal{D}A_\mu][\mathcal{D}\Psi][\mathcal{D}\bar{\Psi}]~\exp\left(-\int_V d^3x \int_0^{1/T} d\tau ~\mathcal{L}_{QCD}\right) 
\end{equation}
We can extract various thermodynamics quantities from $\ln Z(T,V)$ such
as the grand canonical potential ($\Omega(T,V)$), pressure ($p$), or energy
density ($\epsilon$):
\begin{equation}
\Omega(T,V)  =  T \ln Z(T,V); ~~\epsilon = \frac{E}{V} = -\frac{1}{V}\frac{\partial \ln Z}{\partial (1/T)};~~ p = T \frac{\partial \ln Z}{\partial V}
\end{equation}
In the thermodynamic limit, the grand canonical potential is an
extensive quantity ($\Omega \sim V$).  Thus, we can write:
\begin{equation}
p = \frac{T}{V} \ln Z(T,V); ~~ \epsilon = \frac{T^2}{V} \frac{\partial \ln Z(V,T)}{\partial T}
\end{equation}
We can also define another quantity, called the interaction measure $I$, 
which is the trace of the energy-momentum tensor:
\begin{equation}
\label{eqn:interaction}
\frac{I}{T^4} = \frac{\epsilon - 3p}{T^4} = T\frac{\partial}{\partial T}(\frac{p}{T^4})
\end{equation}

Unfortunately, calculating the pressure by computing $\ln Z(T,V)$ directly
is not possible using stochastic Monte Carlo methods.  However, we are able to
compute certain derivatives of $\ln Z(T,V)$ which, in the lattice formulation, become
the expectation values of operators.  In particular, we can obtain the
gauge action and the chiral condensate in this manner:
\begin{equation}
\left<S_g\right> = \frac{\partial \ln Z}{\partial \beta};~~ \left<\bar{\psi}\psi_q\right> = \frac{\partial \ln Z}{\partial \tilde{m}_q}
\end{equation}
where $\beta = 6/g_0^2$ is related to the bare coupling $g_0$ and $\tilde{m}_q$
denotes a bare quark mass.

On the lattice, finite temperatures can be simulated
by limiting the temporal size of the lattice.  This is directly analagous
to restricting the integral in \ref{eqn:partition} to a finite interval
in the imaginary time direction.
The temperature is related to the temporal extent by $T^{-1} = N_t a(g_0, 
\tilde{m}_q(g_0))$, where the lattice spacing $a$ is a function of
both $g_0$ and $\tilde{m}_q$, and the bare quark masses $\tilde{m}_q$ must
be adjusted as a function of $g_0$ along some renormalization group trajectory
to keep physical quantities fixed, i.e. a line of constant physics.

We can rewrite the expression \ref{eqn:interaction} for the 
interaction measure in terms of lattice quantities:
\begin{equation}
\label{eqn:interaction2}
\frac{I}{T^4} = \left(\frac{N_t}{N_s}\right)^3 \left(\frac{\partial \beta}{\partial\ln a}\left(\left<S_g(T)\right>-\left<S_g(0)\right>\right) + \sum_q \frac{\partial \tilde{m}_q}{\partial\ln a}\left(\left<\bar{\psi}\psi_q(T)\right>-\left<\bar{\psi}\psi_q(0)\right>\right)\right)
\end{equation}
Here, the sum over the index $q$ indicates a sum over all quark flavors.  We
have also normalized the interaction measure by subtracting the $T=0$
contribution, disentangling the vacuum contribution from the thermal effects.

Using $I$, we can reconstruct other thermodynamic quantities.  For example,
we use \ref{eqn:interaction} to express the pressure in terms of lattice
quantities:
\begin{eqnarray}
\frac{P}{T^4} & = & \int_0^T \frac{I}{T'^4} d \ln T'\\
\frac{P}{T^4} & = & \left(\frac{N_t}{N_s}\right)^3 \int_{\beta_0}^{\beta} \left(\left(\left<S_g(T)\right>-\left<S_g(0)\right>\right) + \sum_q \frac{\partial \tilde{m}_q}{\partial\beta'}\left(\left<\bar{\psi}\psi_q(T)\right>-\left<\bar{\psi}\psi_q(0)\right>\right)\right)~d\beta'
\end{eqnarray}
Once we have $I(T)$ and $p(T)$, we can easily reconstruction the 
energy density, $\epsilon(T)$, or the entropy density, $s(T)$.
For a more detailed discussion of our EoS calculation with
2+1 flavors, see  \cite{Cheng:2007tp}.

\section{Calculation Method}
For this calculation, we are concerned only with the charm quark contribution.
The relevant part of \ref{eqn:interaction2} is:
\begin{equation}
\frac{I_c}{T^4} = \left(\frac{N_t}{N_s}\right)^3 \frac{d\beta}{d \ln a}\frac{\partial \tilde{m}_c}{\partial\beta}\left(\left<\bar{\psi}\psi_c(T)\right>-\left<\bar{\psi}\psi_c(0)\right>\right)
\end{equation}

To calculate $\bar{\psi}\psi_c$, we make partially quenched measurements of
the chiral condensate on previously generated 2+1f gauge configurations,
both at finite temperature ($N_t = 4$) and zero temperature ($N_t = 32$).
These dynamical configurations use the p4fat3 fermion action \cite{Karsch:2000ps}
, and a tree-level improved Symanzik gauge action.

In order to determine the temperature of each ensemble, we have chosen to set 
the scale using the static quark potential parameter $r_0$.  
$r_0$ is defined as:
\begin{equation}
\left(r^2\frac{d V_{q\bar{q}}(r)}{dr}\right)_{r=r_0} = 1.65
\end{equation}
We can then convert to physical units by using $r_0 = 0.469(7)$ fm.
 \cite{Gray:2005ur}.
Note that the bare quark masses ($\tilde{m}_{ud}$ and $\tilde{m_s}$) are
tuned as a function of $\beta$ so that physical quantities such as 
$m_{\pi}r_0$, $m_{\eta}r_0$, and $m_K r_0$ take on approximately constant 
values over the entire temperature range.  

To determine $I_c$, we not only need $\bar{\psi}\psi_c$ on zero and finite
temperature lattices, but also $d \beta/d \ln a$.  We can
deduce $d \beta/d \ln a$ from our scale-setting calculations of $r_0$:
\begin{equation}
\frac{d \beta}{d \ln a} = a \frac{d \beta}{d a} = \left(\frac{\partial \ln (a/r_0)}{\partial \beta}\right)^{-1}
\end{equation}
To get a smooth function for $d \beta/d \ln a$, we fit $\frac{a}{r_0}(\beta)$
to a renormalization-group inspired ansatz.  Table 1 gives details of
the input parameters for the lattices, the number of trajectories for
the different ensembles, as well as $r_0$ and $T/T_c$.  For more details about
scale setting and the line of constant physics, see  \cite{Cheng:2007tp}.

\begin{table}[hbt]
\begin{center}
\begin{tabular}{|cccccccrr|}
\hline
& & & & & & & \multicolumn{2}{c|}{Trajectories}\\             
$\beta$ & Volume & $\tilde{m}_{ud}$ & $r_0/a$ & $T/T_c$ & $\tilde{m}_c (\eta_c)$ & $\tilde{m}_c (J/\Psi)$ & $N_t = 32$ & $N_t = 4$ \\
\hline
3.277 & $16^3$ & .00765  & 1.797(19) & 0.90 & 2.37 & 2.25 & 3250 & 12160\\
3.335 & $16^3$ & .00570  & 2.033(17) & 1.06 & 1.57 & 1.42 & 2630 & 14280\\
3.351 & $16^3$ & .00592  & 2.069(12) & 1.10 & 1.50 & 1.35 & 6950& 12420\\
3.382 & $16^3$ & .00520  & 2.225(13) & 1.20 & 1.22 & 1.11 & 2270& 8110\\
3.41 & $16^3$  & .00412  & 2.503(18) & 1.31 & 1.02 & .858 & 2790& 16000\\
3.46 & $16^3$  & .00313  & 2.890(16) & 1.50 & .670 & .650 & 2510& 10200\\
3.49 & $16^3$  & .00290  & 3.223(31) & 1.62 & .566 & .529 & 4290& 9420\\
3.51 & $16^3$  & .00259  & 3.423(61) & 1.70 & .508 & .473 & 2450& 10000\\
3.54 & $16^3$  & .00240  & 3.687(34) & 1.83 & .446 & .417 & 4060& 6250\\
3.57 & $24^3$  & .00212  & 4.009(26) & 1.98 & .386 & .347 & 2460& 21190\\
3.63 & $24^3$  & .00170  & 4.651(41) & 2.28 & .304 & .288 & 3290& 10000\\
3.69 & $24^3$  & .00150  & 5.201(48) & 2.61 & .257 & .244 & 2290& 9470\\
3.76 & $24^3$  & .00130  & 6.050(61) & 3.05 & .213 & .205 & 1110& 33370\\
3.82 & $24^3$  & .00110  & 6.752(96) & 3.46 & .181 & .190 & 3000& 35000\\
3.92 & $24^3$  & .00092  & 7.59(12)  & 4.23 & .162 & .157 & 4080& 35870\\
\hline
\end{tabular}
\end{center}
\label{tab:parameters}
\caption{Input parameters for the ensembles on which we have performed
measurements.  Note, $\tilde{m}_s = 10 \tilde{m}_{ud}$ for all of these
ensembles.  Also given are $r_0$, $T/T_c$, $\tilde{m}_c$ determined from
$m_{\eta_c}$ and $m_{J/\Psi}$, and the number of trajectories for $N_t = 4$ and $N_t = 32$.}
\vspace{-.2cm}
\end{table}

\section{Setting the physical charm mass}
\label{sec:mc}
In addition to $d \beta/d \ln a$, we also need $\partial \tilde{m}_c/\partial \beta$ 
to calculate $I_c$.  Thus, we need some method to determine the
bare charm quark mass so that it also sits on a line of constant physics as
we vary $\beta$.

Ideally, if we were simulating in the scaling regime for charm quark
observables, the ratio $\tilde{m}_c/\tilde{m}_{ud}$ should
be a constant throughout the entire temperature range and independent of
the quantity chosen to fix $\tilde{m}_c$.  Unfortunately, 
$\tilde{m}_c$ is quite large on the available lattices, so the ratio
$\tilde{m}_c/\tilde{m}_{ud}$ depends on $\beta$ and the particular quantity
used to set $\tilde{m}_c$.

\begin{figure}[t]
\includegraphics[width=0.50\textwidth]{mcratio.eps}
\includegraphics[width=0.50\textwidth]{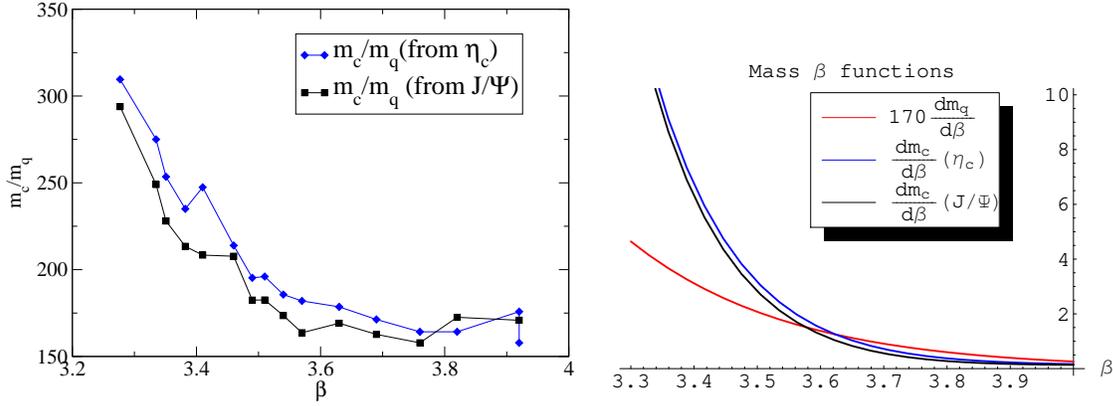}
\caption{On the left, we see $\tilde{m}_c/\tilde{m}_{ud}$ as a function of
$\beta$.  On the right, we see $\partial \tilde{m}_c/\partial \beta$.}
\label{fig:mc}
\vspace{-.2cm}
\end{figure}

In order to understand the extent of this scaling violation, and how it affects
our calculation, we have chosen to fix $\tilde{m}_c$ by setting two
different charmonium states  to their physical masses ($m_{\eta_c}$ = 
2.980 GeV and $m_{J/\Psi}$ = 3.097 GeV).
This produces a range of values $0.16 < \tilde{m}_c < 2.4$, with exact figures
given in Table 1.  As seen in Figure \ref{fig:mc}, we find fairly reasonable 
scaling ($\tilde{m}_c/\tilde{m}_{ud} \sim 170$) for the finer lattice spacings 
($\beta > 3.5 \rightarrow T > 1.7 T_c$).  
However, as we lower $\beta$ onto coarser lattices,
we find that this approximate scaling breaks down, and this ratio
$\tilde{m}_c/\tilde{m}_{ud}$ begins to increase dramatically.  We also find 
that using the $\eta_c$ state tends to give a systematically higher value for 
$\tilde{m}_c$, although this ultimately does not have a large affect on the 
final calculation.

Once we have $\tilde{m}_c(\beta)$, we can fit this data to a RG-inspired ansatz
for the running of the bare quark mass.  This allows us to calculate a smooth 
curve for $\partial \tilde{m}_c/\partial \beta$, also shown in Figure 
\ref{fig:mc}.  We see that this $\beta$-function increase s
dramatically as we move to smaller $\beta$, in contrast to the scaled version 
of $\partial \tilde{m}_{ud}/\partial \beta$, which shows a much milder 
$\beta$-dependence.

\section{Results}

\begin{figure}[hbt]
\vspace{-1cm}
\includegraphics[width=0.50\textwidth]{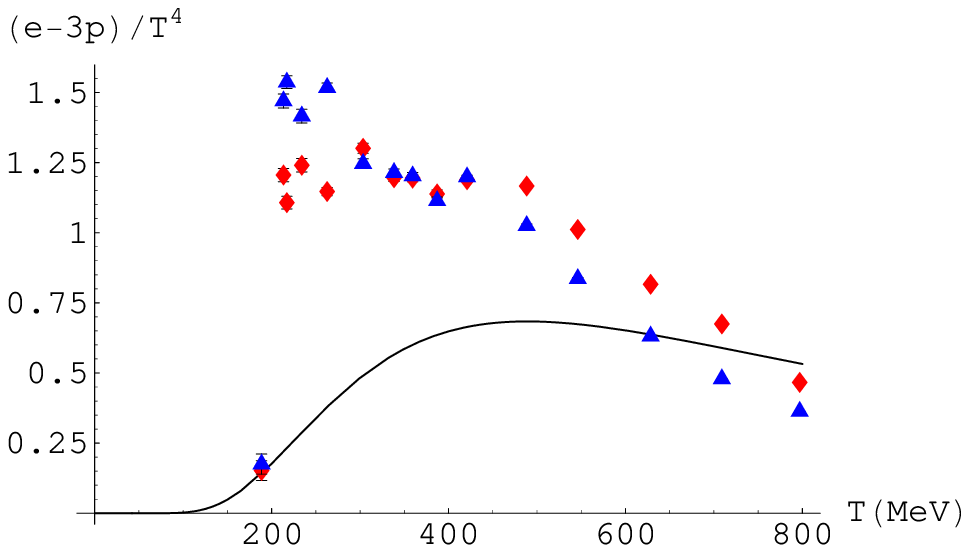}
\includegraphics[width=0.50\textwidth]{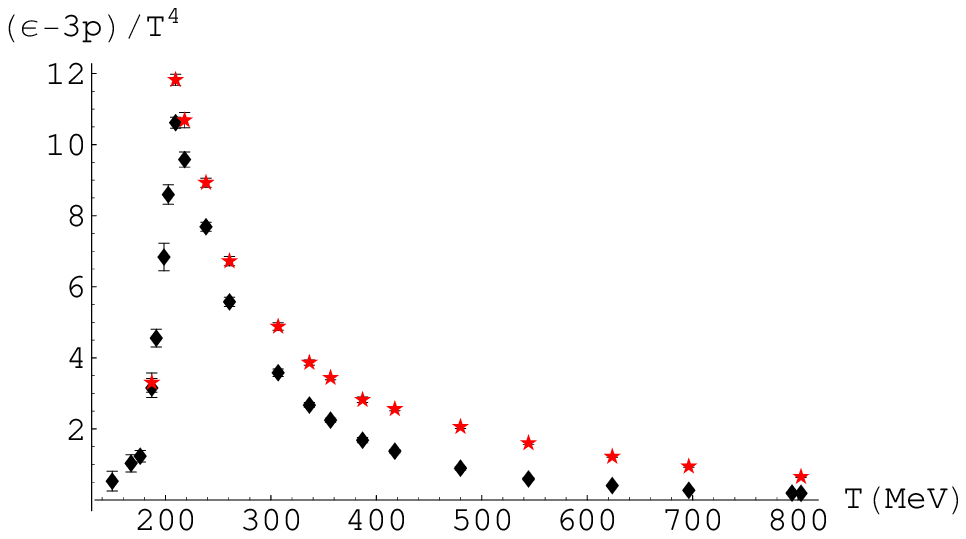}
\includegraphics[width=0.50\textwidth]{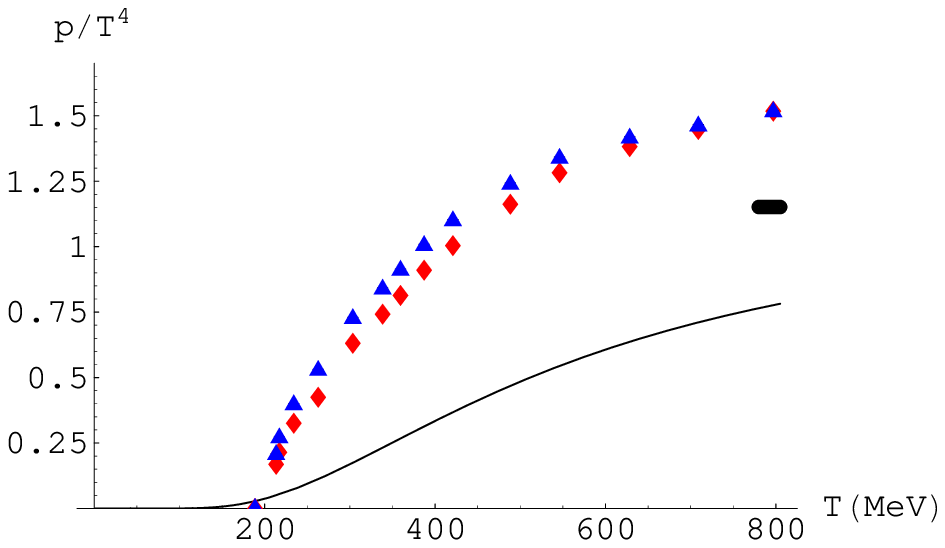}
\includegraphics[width=0.50\textwidth]{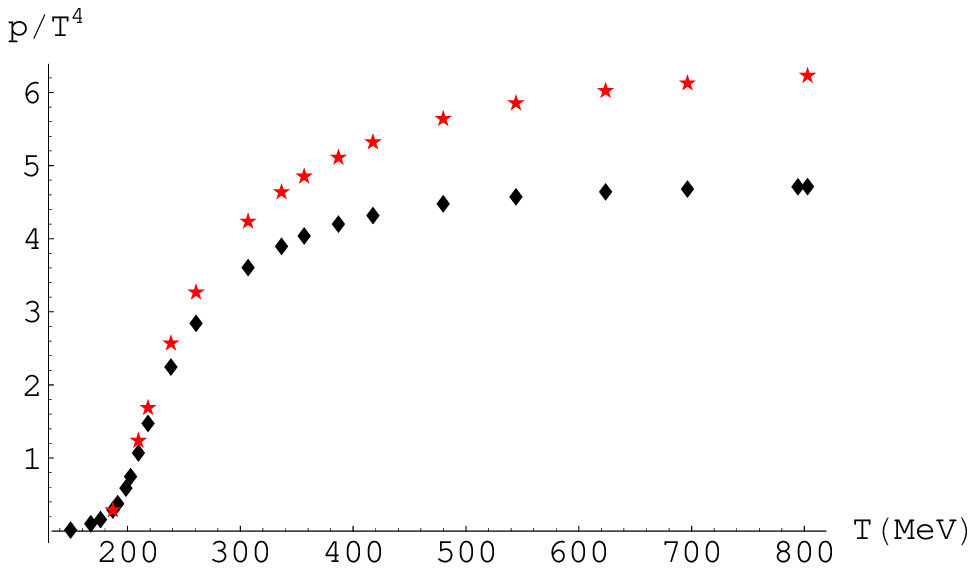}
\includegraphics[width=0.50\textwidth]{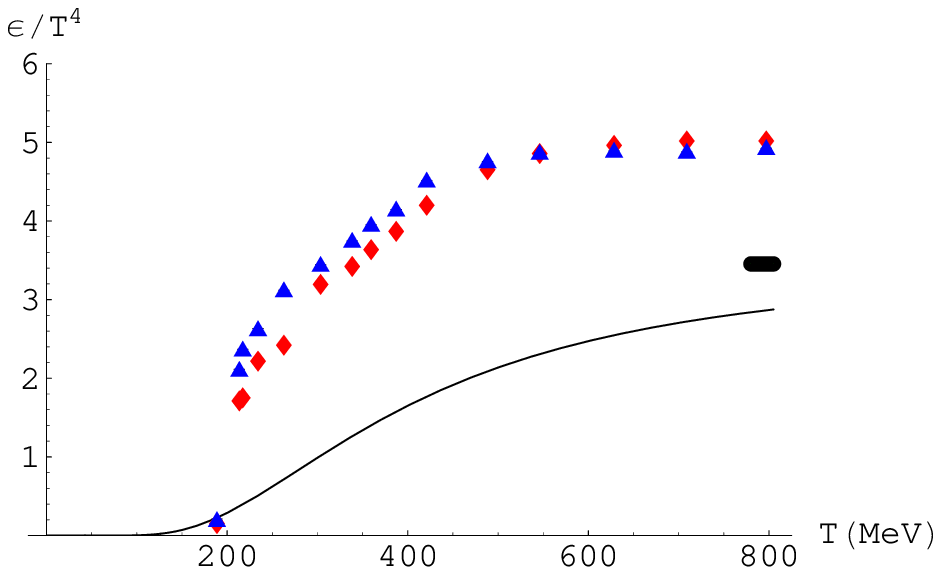}
\includegraphics[width=0.50\textwidth]{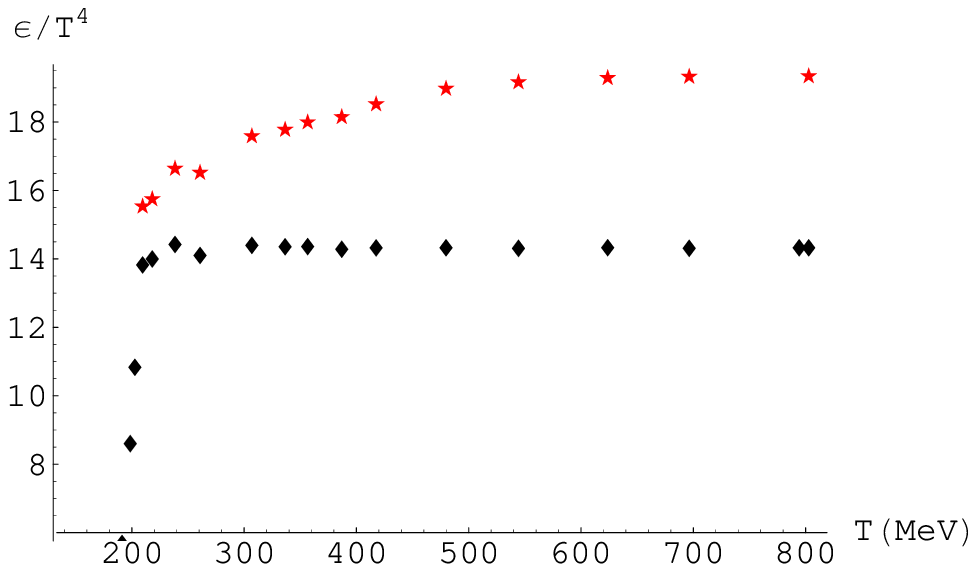}
\caption{In the left column, the charm contribution 
to the interaction measure ($\frac{\epsilon-3p}{T^4}$), 
pressure ($\frac{p}{T^4}$), and energy density ($\frac{\epsilon}{T^4}$), 
respectively.  
Diamonds use $\eta_c$ while the triangles use $J/\Psi$ to set 
$\tilde{m}_c$.  The solid curve gives the free-field results, while the solid bars 
on the upper right give the $T\rightarrow \infty$ limit.  In the right column, the
charm contribution (stars) is added onto the pure 2+1f part (diamonds).}
\label{fig:results}
\vspace{-0.3cm}
\end{figure}

Figure \ref{fig:results} shows our results for the charm contribution using
both $\eta_c$ and $J/\Psi$ to determine $\tilde{m}_c$, as well as a free-field
calculation using the physical value of the charm mass $m_c \approx 1.2 GeV$.
The choice of $\eta_c$ or $J/\Psi$ does have some effect, but both curves
share the same qualitative features.  Notably, the interaction measure $I_c$
increases drastically as we move through the transition.  This is consistent
with the notion that the charmonium states are the lightest charm states
that contribute below the transition, and thus are exponentially suppressed
by their heavy mass.  

It is interesting to note that the value of $I_c$ already becomes quite
large even just above the transition ($T \sim 1.1 T_c$), decaying away slowly
as the temperature is increased.  This is in contrast to the free-field 
calculation, where the interaction measure peaks at $T \sim 0.35 m_c \approx 
400 MeV$, although of course the free-field result knows nothing about the
crossover transition.  Perhaps more surprising is the fact that $I_c$ 
is significantly greater than the free-field value for $T < 3 T_c$.  
As a result, $p$ and $\epsilon$ increase much faster just
above the transition than expected.  This causes $p$ and
$\epsilon$ to overshoot their continuum Stefan-Boltzmann values $T \rightarrow 
\infty$ by a large amount, even at finite temperature.  This difference 
may be attributed to finite lattice spacing corrections at $N_t = 4$.  Indeed,
for the 2+1f calculation, the contribution of the quark condensates to 
$(\epsilon-3p)/T^4$ decreases going from $N_t = 4$ to $N_t=6$.

Figure \ref{fig:results} also
 shows the charm contribution superimposed on the 2+1f
calculation at $N_t = 4$.  Although we see little change in the interaction
measure because of the high 2+1f peak, the pressure and energy change by
quite a large amount for $T > T_c$.

\section{Conclusion}
We have made a partially quenched calculation of the charm quark equation of
state.  Current EoS studies are in a temperature regime that is a significant
fraction of the charm quark mass, where the dynamics of the charm quark
may have a significant effect and can no longer be ignored.

Admittedly, this calculation has unquantified, possibly large systematic
errors.  First of all, our calculations are partially quenched, so that
the full dynamics of the charm quark may not be accurately reflected in their
"back-reaction" on the gauge fields.  To the extent that the charm quark is heavy,
this should be a small effect, but may be problematic at higher temperatures and
at finer lattice spacings.
Secondly, $\tilde{m}_c$ is a large fraction of, or even exceeds, the lattice
spacing on many of the lattices on which we measure.  This introduces large
cut-off effects, and it is appropriate to ask whether the p4 fermion
formulation can give even a reasonable facsimile of the charm quark dynamics at
these heavy masses.

As there are no plans for large-scale simulations involving a dynamical charm
quark, it seems we must tolerate the first problem.  The second problem
may be alleviated somewhat be moving to finer lattice spacings, 
($N_t = 6, 8$).  This may give us a better grasp of the discretization errors
in the current calculation.  
Furthermore, the recent development of the
HISQ action \cite{Follana:2006rc}, which, among other improvements, removes the leading order 
$O(ma)^4$ corrections to the dispersion relation at tree-level, provides
another possible tool to mitigate the large cut-off effects currently present.

The future plans for this calculation is extension to $N_t = 6$ and possibly
$N_t = 8$, as well as investigating the possibility of adapting the methods
used by the HISQ action.  Hopefully, this will allow us to better understand
why the $N_t = 4$ calculation so badly overshoots the Stefan-Boltzmann limit,
and to better quantify the cut-off effects.

\vspace{-.2cm}
\section{Acknowledgments}
This work was carried out in collaboration with the RBC-Bielefeld 
Collaboration.  In particular we thank Norman Christ, Frithjof Karsch, and 
Peter Petreczky for useful discussions.  Computations were performed on the 
RIKEN-BNL QCDOC machines at Brookhaven National Laboratory.  We thank 
RIKEN, BNL, Columbia University, and the US DOE for providing the facilities 
on which this work was done. This research was supported by US DOE grants 
DE-AC02-98CH1-886 and DE-FG02-92ER40699.

\end{document}